\documentclass[twocolumn]{aastex631}
\usepackage{apjfonts}
\usepackage{graphicx}
\usepackage{amsmath}
\usepackage{amssymb}
\usepackage{color}

\gdef\xx[#1]{\textcolor{red}{#1}}

\gdef\kms{km\,s$^{-1}$}
\gdef\msun{M$_{\odot}$}

\newcommand{\GG}[1]{}
\lefthead{van Dokkum et al.}
\righthead{}
\begin{document}

\newcommand\XXX[1]{{\textcolor{red}{\textbf{x\ #1\ x}}}}

%\title{Multiple candidate
%runaway supermassive black holes associated with
%a galaxy at $z=0.96$}

\title{A candidate
runaway supermassive black hole identified by shocks
and star formation in its wake}

%\correspondingauthor{Pieter van Dokkum}

\author[0000-0002-8282-9888]{Pieter van Dokkum}
\affiliation{Astronomy Department, Yale University, 52 Hillhouse Ave,
New Haven, CT 06511, USA}

\author[0000-0002-7075-9931]{Imad Pasha}
\affiliation{Astronomy Department, Yale University, 52 Hillhouse Ave,
New Haven, CT 06511, USA}

\author[0000-0003-3153-8543]{Maria Luisa Buzzo}
\affiliation{Swinburne University of Technology, Melbourne, Victoria, Australia}

\author[0000-0002-5907-3330]{Stephanie LaMassa}
\affiliation{Space Telescope Science Institute, 3700 San Martin Drive,
Baltimore, MD 21218, USA}

\author[0000-0002-5120-1684]{Zili Shen}
\affiliation{Astronomy Department, Yale University, 52 Hillhouse Ave,
New Haven, CT 06511, USA}

\author[0000-0002-7743-2501]{Michael A.\ Keim}
\affiliation{Astronomy Department, Yale University, 52 Hillhouse Ave,
New Haven, CT 06511, USA}

\author[0000-0002-4542-921X]{Roberto Abraham}
\affiliation{Department of Astronomy \& Astrophysics, University of Toronto,
50 St.\ George Street, Toronto, ON M5S 3H4, Canada}

\author[0000-0002-1590-8551]{Charlie Conroy}
\affiliation{Harvard-Smithsonian Center for Astrophysics, 60 Garden Street,
Cambridge, MA, USA}

\author[0000-0002-1841-2252]{Shany Danieli}
\altaffiliation{NASA Hubble Fellow}
\affiliation{Department of Astrophysical Sciences, 4 Ivy Lane, Princeton University, Princeton, NJ 08544, USA}

\author[0000-0001-8073-4554]{Kaustav Mitra}
\affiliation{Astronomy Department, Yale University, 52 Hillhouse Ave,
New Haven, CT 06511, USA}

\author[0000-0002-6766-5942]{Daisuke Nagai}
\affiliation{Department of Physics, Yale University, P.O. Box 208121, New Haven, CT 06520, USA}

\author[0000-0002-5554-8896]{Priyamvada Natarajan}
\affiliation{Astronomy Department, Yale University, 52 Hillhouse Ave,
New Haven, CT 06511, USA}

\author[0000-0003-2473-0369]{Aaron J.\ Romanowsky}
\affiliation{Department of Physics and Astronomy, San Jos\'e State University,
San Jose, CA 95192, USA}
\affiliation{Department of
Astronomy and Astrophysics, University of California Santa Cruz, 1156 High Street, Santa Cruz, CA 95064, USA}

\author[0000-0002-5445-5401]{Grant Tremblay}
\affiliation{Harvard-Smithsonian Center for Astrophysics, 60 Garden Street,
Cambridge, MA, USA}

\author[0000-0002-0745-9792]{C.\ Megan Urry}
\affiliation{Department of Physics, Yale University, P.O. Box 208121, New Haven, CT 06520, USA}

\author[0000-0003-3236-2068]{Frank C.\ van den Bosch}
\affiliation{Astronomy Department, Yale University, 52 Hillhouse Ave,
New Haven, CT 06511, USA}

\begin{abstract}

The interaction of a runaway supermassive black hole (SMBH)
with the circumgalactic medium (CGM)
can lead to the formation of a wake of shocked gas and young stars behind it.
Here we report the serendipitous
discovery of  an extremely narrow linear feature in HST/ACS images that may
be an example of such a wake. The feature
extends 62\,kpc
from the nucleus of a compact star-forming galaxy at $z=0.964$.
Keck LRIS spectra show that the
[O\,{\sc iii}]/H$\beta$
ratio varies from $\sim 1$ to $\sim 10$ along the
feature, indicating a
mixture of star formation and fast shocks.
The feature 
terminates in a bright [O\,{\sc iii}] knot with a luminosity
of $\approx 1.9 \times 10^{41}$\,ergs\,s$^{-1}$.
The stellar continuum colors vary along the feature, and
are well-fit by a simple model that has a
monotonically increasing age with distance from the tip.
The line ratios, colors, and the overall morphology are
consistent with an ejected SMBH moving through the CGM at high speed
while triggering star formation. The best-fit
time since ejection 
is $\sim 39$\,Myr and the implied velocity
is $v_{\rm BH}\sim 1600$\,\kms. 
The feature is not perfectly straight in the HST
images, and we show that the amplitude of the
observed spatial variations is consistent with the
runaway SMBH interpretation.
%If confirmed, this
%would be the first of a possibly large population of runaway SMBHs.
Opposite the primary wake is a fainter and shorter
feature, marginally detected in
[O\,{\sc iii}] and the rest-frame far-ultraviolet. This feature
may be shocked gas behind
a binary SMBH that was ejected at the same time as the SMBH that
produced the primary wake.

\end{abstract}

%\keywords{
%galaxies: evolution --- galaxies: structure }

\section{Introduction}

There are several ways for a supermassive black hole (SMBH) to escape
from the center of a galaxy. The first step is always a galaxy merger,
which leads to the formation of a binary SMBH at the center of the
merger remnant \citep{begelman:80,milosavljevic:01}. The binary
can be long-lived, of order $\sim 10^9$\,yr, and if a third SMBH
reaches the center of the galaxy before the binary merges, a three-body
interaction can impart a large velocity to one of the SMBHs leading
to its escape from the nucleus
\citep{saslaw:74,volonteri:03,hoffman:07}. Even in the absence of a third SMBH,
the eventual merger of the binary can impart a kick to the newly formed
black hole through gravitational radiation recoil \citep{bekenstein:73,
campanelli:07}. The velocity of the ejected
SMBH depends on the mechanism and the
specific dynamics. Generally the kicks are expected to be higher for
slingshot scenarios than for recoils \citep[see, e.g.,][]{hoffman:07,kesden:10},
although in exceptional cases recoils may reach $\sim 5000$\,\kms\ 
\citep{campanelli:07,lousto:11}. In
both scenarios the velocity of the SMBH may 
exceed the escape velocity of the host galaxy
\citep[see, e.g.,][]{saslaw:74,hoffman:07,lousto:12,ricarte:21}.

Identifying such runaway SMBHs is of obvious interest but difficult. The main focus has been on the special case where
the black hole is accreting at a high enough rate to be identified as a
kinematically
or spatially displaced active galactic nucleus (AGN) \citep{bonning:07,
blecha:11,komossa:12}. For such objects, the presence of a SMBH is not
in doubt, but  it can be difficult to determine whether they are
``naked'' black holes or the nuclei of merging galaxies
\cite[see, e.g.,][]{merritt:06}.
Candidates include the peculiar double X-ray source
CID-42 in the COSMOS field \citep{civano:10} and the
quasars HE0450--2958 \citep{magain:05},
SDSSJ0927+2943 \citep{komossa:08}, E1821+643
\citep{robinson:10,jadhav:21}, and
3C\,186 \citep{chiaberge:17}.

Quiescent (non-accreting) runaway
SMBHs
can be detected through the effect they have on their
surroundings.
As noted by \citet{boylankolchin:04} and discussed
in-depth by \citet{merritt:09},
some of the stars in the nuclear regions of the galaxy
are expected to remain bound to the SMBH during and after its
departure. The
stellar mass that accompanies the
black hole is a steeply declining function of its velocity,
and generally $\lesssim M_{\rm BH}$. This leads
to peculiar objects, dubbed ``hyper compact stellar systems'' (HCSS) by
\citet{merritt:09}, with the sizes and luminosities of globular clusters
or ultra compact dwarf galaxies but the velocity dispersions of
massive galaxy nuclei. HCSSs could therefore be easily identified
by their kinematics, but
measuring velocity dispersions of such faint
objects is difficult beyond the very local Universe.
Other potential
detection methods include gravitational lensing \citep{sahu:22}
and tidal disruption events
\citep[e.g.,][]{ricarte:21b,angus:22}.
No convincing candidates have been found so far.

Another way to identify runaway
SMBHs is through the effect of their passage
on the surrounding gas. This topic has an interesting history as it is
rooted in AGN models that turned out to be dead ends.
\citet{saslaw:72} investigated
the suggestion by \citet{burbidge:71} and \citet{arp:72} that the
redshifts of quasars are not cosmological but that they were
ejected from nearby galaxies. In that context they studied what
happens when a SMBH travels  supersonically through
ionized hydrogen, finding that this produces a shock
front with a long wake behind it. Shocked
gas clouds in the wake can cool and form stars, potentially illuminating
the wake with ionizing radiation from O stars.
\citet{rees:75} analyzed the possibility that double radio sources
are produced by the interaction of escaped SMBHs with the
intergalactic gas.
They find that this is plausible from an energetics standpoint, although
now we know that the alternative model, feeding of the lobes
by jets emanating from the nucleus \citep{blandford:74}, is the
correct one.

Perhaps because of these somewhat inauspicious
connections with failed AGN models
there has not been a great deal of follow-up work in this area.
To our knowledge, the only study of the
formation of wakes behind runaway SMBHs in a modern context
is \cite{delafuente:08}, who analyze the gravitational effect of the
passage of a SMBH using the impulse approximation.
They find that the SMBH can impart a velocity of a few to
several tens of \kms\ on nearby gas clouds, and that the gas can then
become unstable to fragmentation and star formation. The outcome is
qualitatively similar to the analysis of \citet{saslaw:72}, in the
sense that, under the right conditions,
star formation can occur along the path of the SMBH.

In this paper we report on the serendipitous discovery of a remarkable
linear feature in HST images that we suggest may represent such a SMBH-induced
wake.  We also identify two candidate hyper-compact stellar
systems, one embedded in the tip of the wake and the other on the opposite side
of the galaxy from which they may have escaped.

\section{A 62\,kpc long linear feature at $z=0.964$}

\subsection{Identification in HST/ACS images}

\begin{figure*}[htbp]
  \begin{center}
  \includegraphics[width=1.0\linewidth]{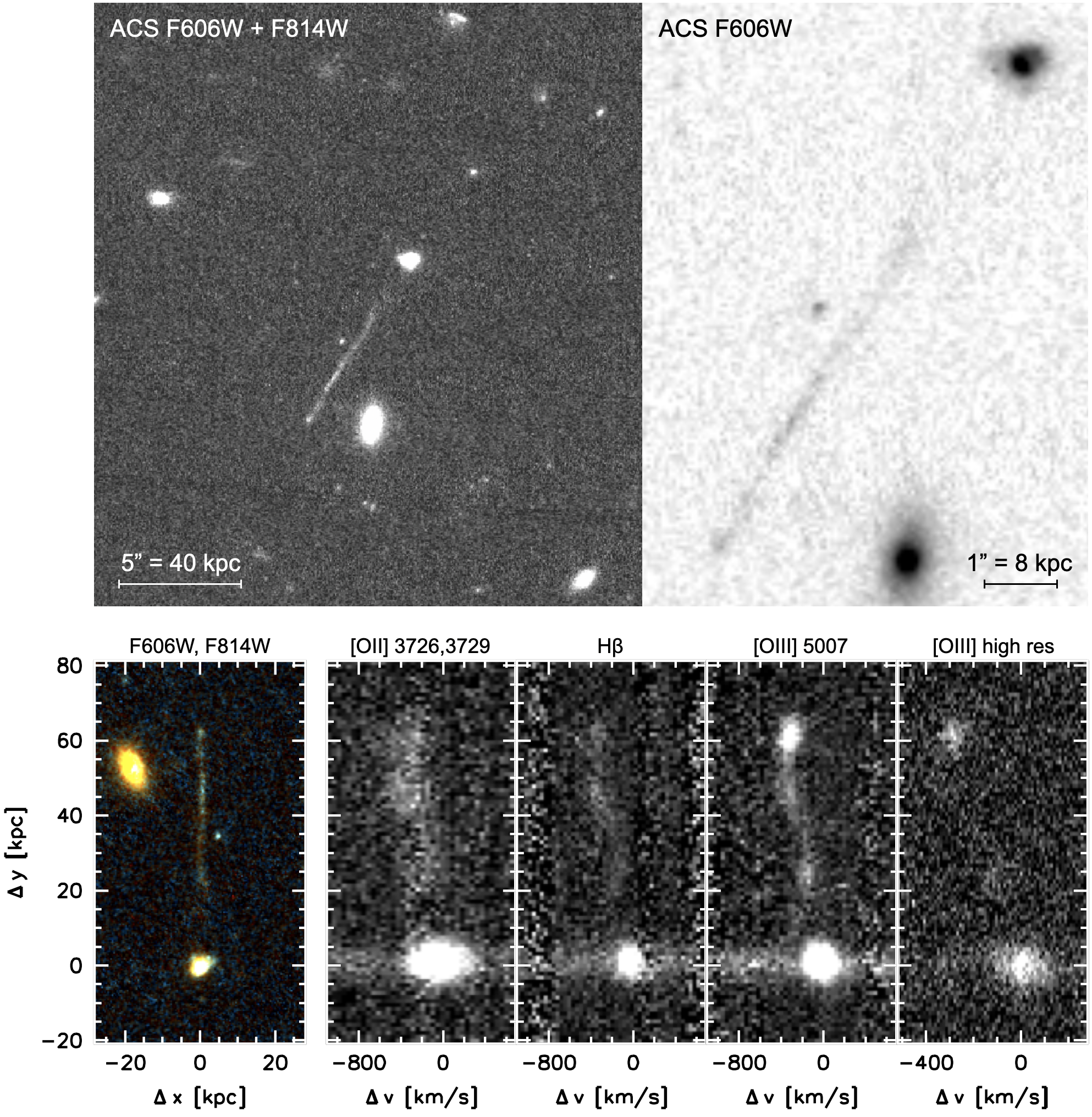}
  \end{center}
\vspace{-0.2cm}
    \caption{
{\em Top left:} F606W\,+\,F814W HST/ACS image of the linear feature and its
surroundings. {\em Top right:} Zoomed view of the F606W image. The
feature shows a compact bright spot at the narrow tip, and seems to broaden
toward the galaxy. {\em Bottom left:} Color image, generated from the
F606W and F814W images. {\em Bottom right panels:} Sections of
LRIS spectra near bright emission lines. The feature and the galaxy
are at the same redshift. The kinematics and line strengths show
complex variations along the feature.
}
\label{overview.fig}
\end{figure*}

We serendipitously identified a thin, linear feature in {\em HST} ACS
images of the nearby dwarf galaxy RCP\,28 \citep{roman:21,dokkum:22},
as shown in Fig.\ \ref{overview.fig}.
RCP\,28 was observed September 5 2022
for one orbit in F606W and one orbit in F814W,
in the context of mid-cycle program GO-16912.
The individual {\tt flc} files were combined using {\tt DrizzlePac}
after applying a flat field correction
to account for drifts in the sensitivity of
the ACS CCDs \citep[see][]{dokkum:22b}. Upon reducing the data
an almost-straight thin streak was readily apparent in
a visual assessment of the data quality (see Fig.\ \ref{overview.fig}).
Based on its appearance we initially thought that it
was a poorly-removed cosmic ray, but the presence of the feature
in both filters quickly ruled out that explanation.
The total AB magnitude of the streak is ${\rm F814W}=22.87 \pm 0.10$
and its luminosity-weighted mean color is ${\rm F606W} - {\rm F814W}
= 0.83\pm 0.05$.

The streak points to the center of a somewhat irregular-looking galaxy,
at $\alpha=2^{\rm h}41^{\rm m}45\fs{}43$; $\delta=-8\arcdeg20\arcmin55\farcs4$
(J2000).  The galaxy has ${\rm F814W}=21.86\pm 0.10$ and 
${\rm F606W} - {\rm F814W} = 0.84 \pm 0.05$; that is, the brightness of the
streak is $\approx 40$\,\% of the brightness of the galaxy and both
objects have the same color within the errors.
Not having encountered something quite like this before in our own
images or in the literature,
we decided to include the feature in the observing plan for
a scheduled Keck run.

\subsection{Redshift}

The feature was observed with the Low-Resolution Imaging
Spectrometer \citep[LRIS;][]{oke:95} on the Keck\,I telescope on
October 1 2022. The 300\,lines\,mm$^{-1}$ grism blazed at 5000\,\AA\
was used on the blue side and the
400\,lines\,mm$^{-1}$ grating blazed at 8500\,\AA\ on the red side,
with the 680\,nm dichroic. The $1\farcs0$ longslit was used, centered on
the galaxy coordinates with a position angle of $327\arcdeg$.
The total exposure time was 1800\,s, split in two exposures of 900\,s.
Conditions were good and the seeing was $\approx 0\farcs 8$.
On October 3 we obtained a high resolution spectrum with the
1200\,lines\,mm$^{-1}$ grating blazed at 9000\,\AA\ in the red. 
Five exposures were obtained for a total exposure time of 2665\,s.
Conditions were highly variable, with fog and clouds hampering the
observations.

Data reduction followed standard procedures for long slit observations.
Sky subtraction and initial wavelength calibration were done with the
{\tt PypeIt} package \citep{pipeit}. The wavelength calibration was
tweaked using sky emission lines, and the data from the individual exposures
were combined. A noise model was created and
cosmic rays were identified as extreme positive deviations from
the expected noise. For the low resolution spectrum
a relative flux calibration, enabling the measurement of
line ratios, was performed using
the spectrophotometric standard HS\,2027.

We find continuum and strong emission lines associated with
the feature. The lines are readily identified
as the redshifted
[O\,{\sc ii}]\,$\lambda\lambda 3726,3729$ doublet, H$\gamma$, H$\beta$, and
[O\,{\sc iii}]\,$\lambda\lambda 4959,5007$. The
redshift is $z=0.964$, and the implied physical extent
of the feature, from the nucleus of the galaxy to its tip, is 62\,kpc. 
The 2D spectrum in the regions around the strongest emission lines is shown in
the bottom panels of Fig.\ \ref{overview.fig}. The lines can be
traced along the entire length of the
feature. There are strong variations in the line strengths and line
ratios, as well as in the line-of-sight velocity. We will return to this in
following sections.
The S/N ratio in the high resolution spectrum is low, about 1/4 of
that in the low resolution spectrum.

\section{Properties of the host galaxy}

\subsection{Morphology}

\begin{figure}[htbp]
  \begin{center}
  \includegraphics[width=0.95\linewidth]{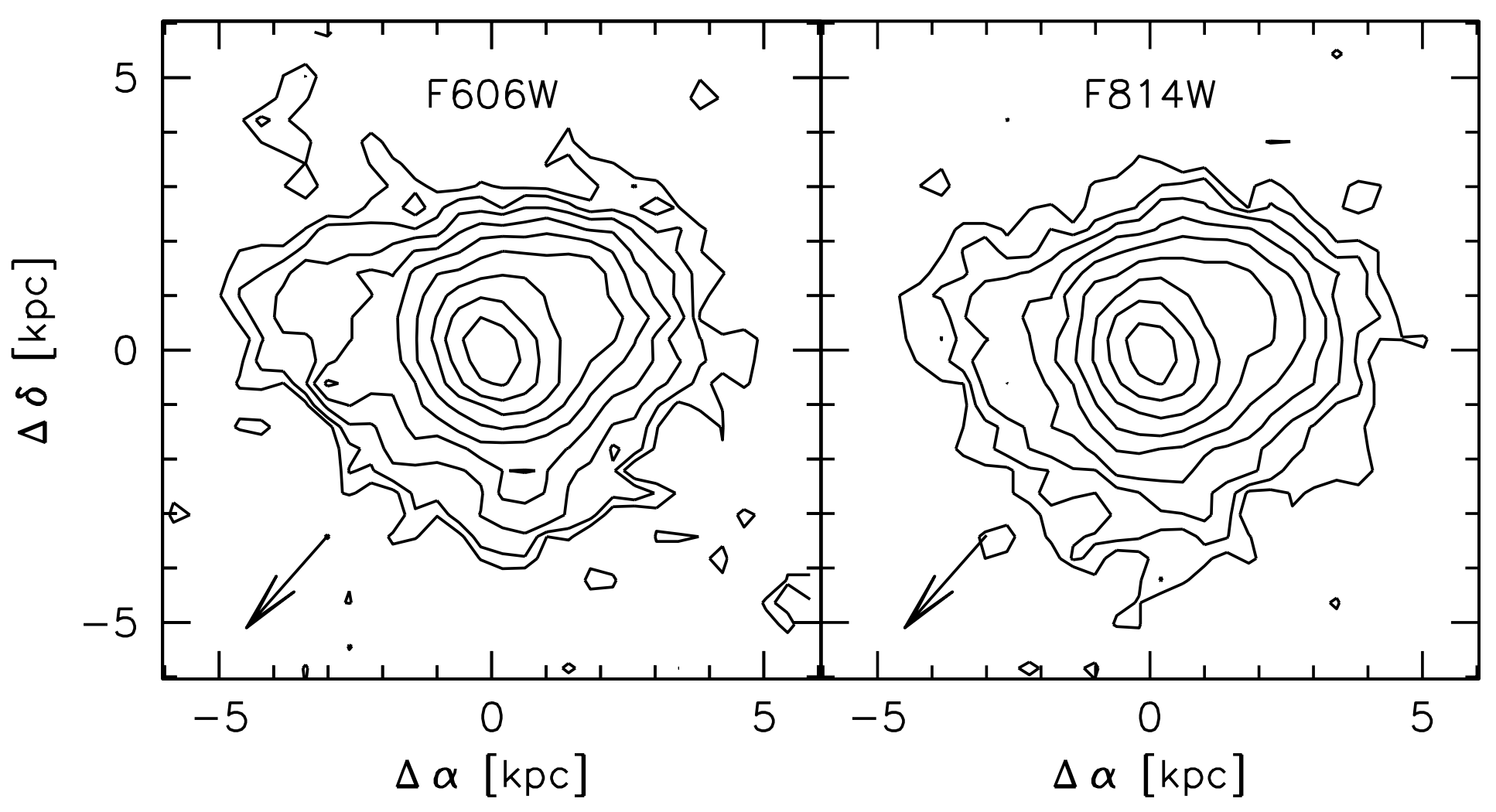}
  \end{center}
\vspace{-0.2cm}
    \caption{
Morphology of the galaxy in F606W and F814W. The arrow indicates the
direction of the linear feature. The galaxy is compact, with
a half-light radius of $r_{\rm e}=1.2$\,kpc, and shows irregular
features possibly indicating a recent merger and/or a connection
to the linear feature.
}
\label{morph.fig}
\end{figure}

The same emission lines are detected in the galaxy, confirming
that it is at the same redshift as the linear feature 
(see Fig.\ \ref{overview.fig}).
The galaxy is compact and somewhat
irregular, as shown in Fig.\ \ref{overview.fig} and by the
contours in Fig.\ \ref{morph.fig}. We determine the half-light
radius of the galaxy with {\tt galfit} \citep{galfit}, fitting
a 2D Sersic profile and using
a star in the image to model the point spread function. 
We find $r_{\rm e} \approx 1.2$\,kpc, but we caution that the
fit has significant residuals. The irregular morphology may
be due to a recent merger or accretion event, although
deeper data are needed to confirm this. 
%Deeper data,
%or data at other wavelengths (see below), can also shed light 
%on the morphological connection
%between the galaxy and the linear feature.

\subsection{Ionization mechanism}

We measure the strength of the strongest emission lines
from the 2D spectra. The continuum was subtracted by fitting a first-order
polynomial in the wavelength direction at all spatial positions, masking
the lines and their immediate vicinity. Line fluxes were
measured by doing aperture photometry on the
residual spectra. 
No corrections for slit
losses or underlying absorption are applied. We find
an [O\,{\sc iii}] flux of
$F = (10\pm 1)\times 10^{-17}$\,ergs\,s$^{-1}$\,cm$^{-2}$
and [O\,{\sc iii}]/H$\beta = 1.9\pm 0.2$.

\begin{figure}[htbp]
  \begin{center}
  \includegraphics[width=0.95\linewidth]{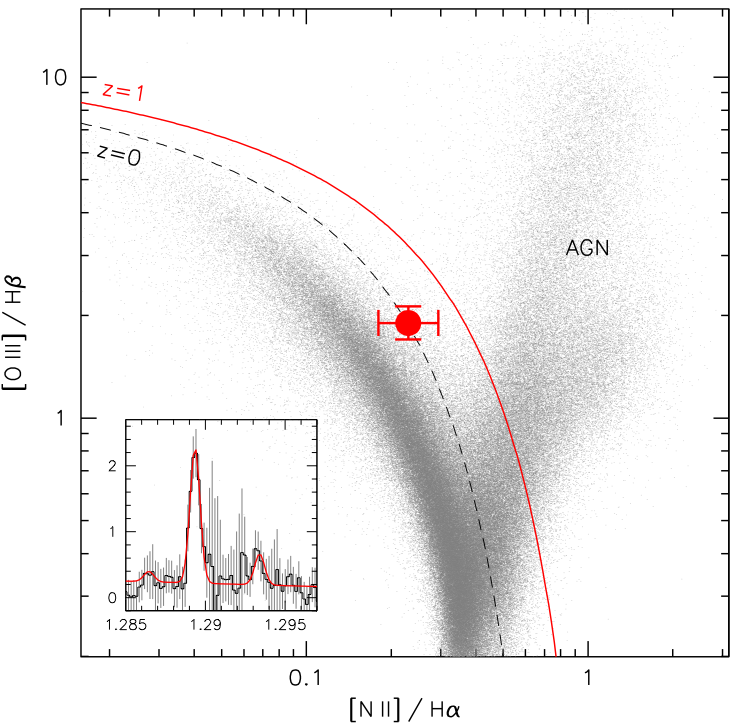}
  \end{center}
\vspace{-0.2cm}
    \caption{
The location of the galaxy in the BPT diagram, with SDSS galaxies
in light grey. The lines divide
``pure'' star forming galaxies from those with an AGN contribution
to their line ratios, for $z=0$ and $z=1$ 
\citep{kewley:13b}. The location is as expected
for a $z=1$ star forming galaxy. The inset shows the
NIRES spectrum in the H$\alpha$ region. The red line is the best fit.
}
\label{bpt.fig}
\end{figure}

The interpretation of the line fluxes depends on the ionization
mechanism, which can be determined from  the combination of
[O\,{\sc iii}]/H$\beta$ and
[N\,{\sc ii}]/H$\alpha$.  H$\alpha$ and [N\,{\sc ii}]
are  redshifted into the $J$ band, and
we observed the galaxy with the Near-Infrared Echellette
Spectrometer (NIRES) on Keck\,II on October 4 2022
to measure these lines.
NIRES provides cross-dispersed near-IR spectra from
$0.94\,\mu$m -- $2.45\,\mu$m
through a fixed $0\farcs 55\times 18\arcsec$ slit. A single 450\,s exposure
was obtained in good conditions, as well as two adjacent empty field exposures.
In the data reduction, the empty field exposures were used for sky subtraction
and sky lines were used for wavelength calibration.
The H$\alpha$ and [N\,{\sc ii}]\,$\lambda 6583$ emission lines of the galaxy
are clearly detected, as shown in the inset of Fig.\ \ref{bpt.fig}.
The emission lines of the galaxy are modeled with the redshift, the
H$\alpha$ line strength, the [N\,{\sc ii}] line strength, and the
velocity dispersion as free parameters. The best-fitting model is
shown in red in Fig.\ \ref{bpt.fig}. We find
a velocity dispersion of $\sigma_{\rm gal} = 60\pm 7$\,\kms\
and [N\,{\sc ii}]/H$\alpha = 0.23 \pm 0.06$, with the uncertainties
determined from bootstrapping.
The implied metallicity, using the \citet{curti:17} calibration,
is $Z=-0.08^{+0.05}_{-0.07}$.

The location of the galaxy in the BPT diagram \citep{baldwin:81}
is shown in Fig.\ \ref{bpt.fig}. For reference,
data from the Sloan Digital Sky Survey (SDSS)
DR7 are shown in grey \citep{brinchmann:04}. The galaxy is
slightly offset from
the SDSS relation of star-forming galaxies and quite far from the AGN
region in the upper right of the diagram. The offset is
consistent with the known changes in the ISM conditions of star
forming galaxies with redshift \citep[see, e.g.,][]{steidel:14,shapley:15}.
The
lines in Fig.\ \ref{bpt.fig}
show the redshift-dependent \citet{kewley:13b} division beyond which AGN
begin to contribute to the line ratios. The galaxy is well within
the ``pure'' star formation region for $z=1$. 

\subsection{Star formation rate and stellar mass}

We infer the star formation rate of the galaxy from the
H$\beta$ luminosity, which is $L_{{\rm H}\beta} = (2.5\pm 0.5) \times
10^{41}$\,ergs\,s$^{-1}$. 
The \citet{kennicutt:98} relation
implies an approximate star formation rate of $6$\,M$_{\odot}$\,yr$^{-1}$
for the dust-free case and $14$\,M$_{\odot}$\,yr$^{-1}$ for
1\,mag of extinction.
The stellar mass of the galaxy can be
estimated from its luminosity and color. We generate
predicted  ${\rm F606W} - {\rm F814W}$ colors
for stellar populations at $z=0.964$ with the {\tt Python-FSPS}
stellar population modeling suite \citep{conroy:09}. We find that
the observed color of the galaxy can be reproduced
with a luminosity-weighted
age of $\sim 150$\,Myr and no dust or an age of $\sim 65$\,Myr
with $A_V\sim 1$. The implied stellar mass is $M_{\rm gal} \sim 7\times
10^9$\,\msun. The typical star formation rate of a galaxy of this
mass at $z=1$ is $\approx 8$\,\msun\,yr$^{-1}$ \citep{whitaker:14},
similar to the observed star formation rate. 

We conclude that the galaxy has
normal  line ratios and a normal
specific star formation rate for its redshift.
Its age is highly uncertain given that
the color is dominated by the most recent star formation, but
if we take the $\sim 100$\,Myr at face value, the past-average
star formation rate is $\sim 70$\,\msun\,yr$^{-1}$, an order
of magnitude larger than the current value.
The galaxy shows morphological irregularities
and is overall quite compact. Its half-light radius of 1.2\,kpc
is a factor of $\sim 3$ smaller than typical galaxies of its
stellar mass and redshift \citep{wel:14}, which implies that
its star formation rate surface density is an order of magnitude
higher. Taken together, these results suggest that the galaxy
experienced a recent merger or accretion event
that led to the funneling of gas into the center and a burst
of star formation $\sim 10^8$\,yr ago.

\section{Shocks and star formation along the feature}

\subsection{Variation in continuum emission and line ratios}

\begin{figure}[htbp]
  \begin{center}
  \includegraphics[width=0.95\linewidth]{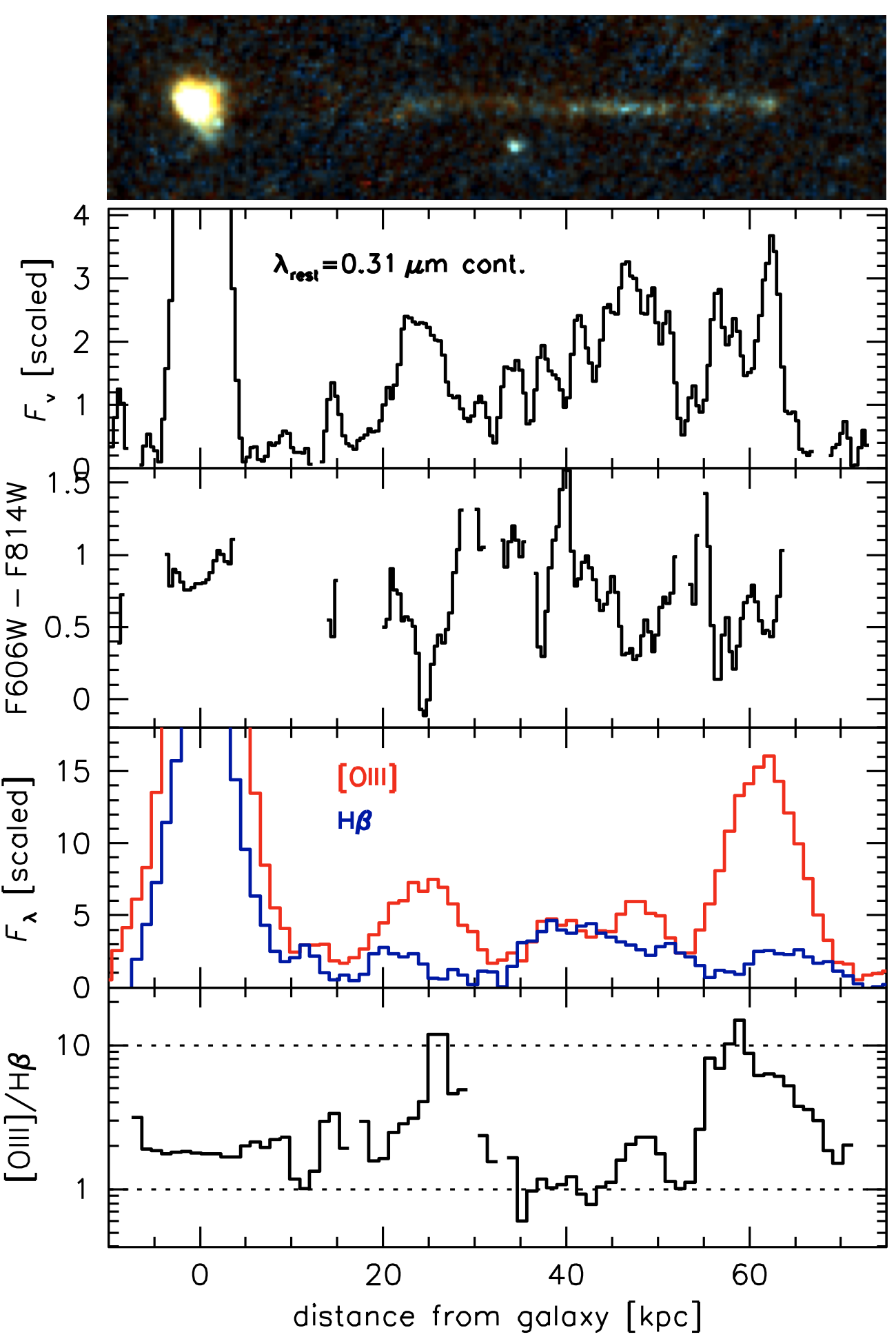}
  \end{center}
\vspace{-0.2cm}
    \caption{
The four panels correspond to
the rest-frame near-UV continuum, ${\rm F606W}-{\rm F814W}$ color,
[O\,{\sc iii}], and H$\beta$ emission
along the linear feature (pictured at the top). The F606W continuum
shows strong variation on all spatial scales, and is brightest at the
furthest point from the galaxy. The color shows large and
seemingly random variations.
The [O\,{\sc iii}]/H$\beta$ ratio 
varies by a factor of $\sim 10$ along the feature,  with some
regions likely dominated 
by shock ionization and others
dominated by H\,{\sc ii} regions.
}
\label{lineratios.fig}
\end{figure}

The linear feature is not uniform in either continuum brightness,
color, line strengths, or line ratios.
The variation along the feature in the F606W ($\lambda_{\rm rest} = 0.31\,\mu$m)
continuum, the ${\rm F606W}-{\rm F814W}$ color,
and in the [O\,{\sc iii}] and H$\beta$ lines is shown in Fig.\
\ref{lineratios.fig}. Note that the spatial resolution of the continuum
emission is $\sim 8\times$ higher than that of the line emission.

There is a general trend of the continuum emission becoming brighter
with increasing distance from the galaxy.
The continuum reaches its peak in a compact
knot at the tip; beyond that point the emission abruptly stops.
As shown in Fig.\ \ref{overview.fig}
the continuum knot at the tip coincides with a
luminous [O\,{\sc iii}] knot in the spectrum. The [O\,{\sc iii}]\,$\lambda
5007$ flux of the knot is $F\approx
3.9\times 10^{-17}$\,ergs\,s$^{-1}$\,cm$^{-2}$,
and the luminosity is $L\approx 1.9\times 10^{41}$\,ergs\,s$^{-1}$.
The [O\,{\sc iii}]/H$\beta$ ratio
reaches $\sim 10$ just behind the knot, higher than can be explained by
photoionization in H\,{\sc ii} regions.

The ionization source could be an AGN, although as
discussed in more detail in \S\,\ref{locations.sec}
the [O\,{\sc iii}] emission
is so bright that an
accompanying X-ray detection might be expected in existing
Chandra data. 
An alternative interpretation is that the bright
[O\,{\sc iii}] knot is caused by a strong shock
\citep[see][]{shull:79,dopita:95,allen:08}.
In the models of \citet{allen:08},
photoionization ahead of a fast ($\gtrsim 500$\,\kms)
shock is capable of producing [O\,{\sc iii}]/H$\beta \sim 10$,
and the expected associated soft X-ray emission
\citep{dopita:96,wilson:99} may be below current detection limits.
There is at least one more region
with elevated [O\,{\sc iii}]/H$\beta$ ratios (at $r \approx
25$\,kpc), and the [O\,{\sc iii}] emission
near the tip could simply be the strongest
of a series of fast shocks along the length of the feature.

\subsection{Stellar populations}

\begin{figure}[htbp]
  \begin{center}
  \includegraphics[width=0.95\linewidth]{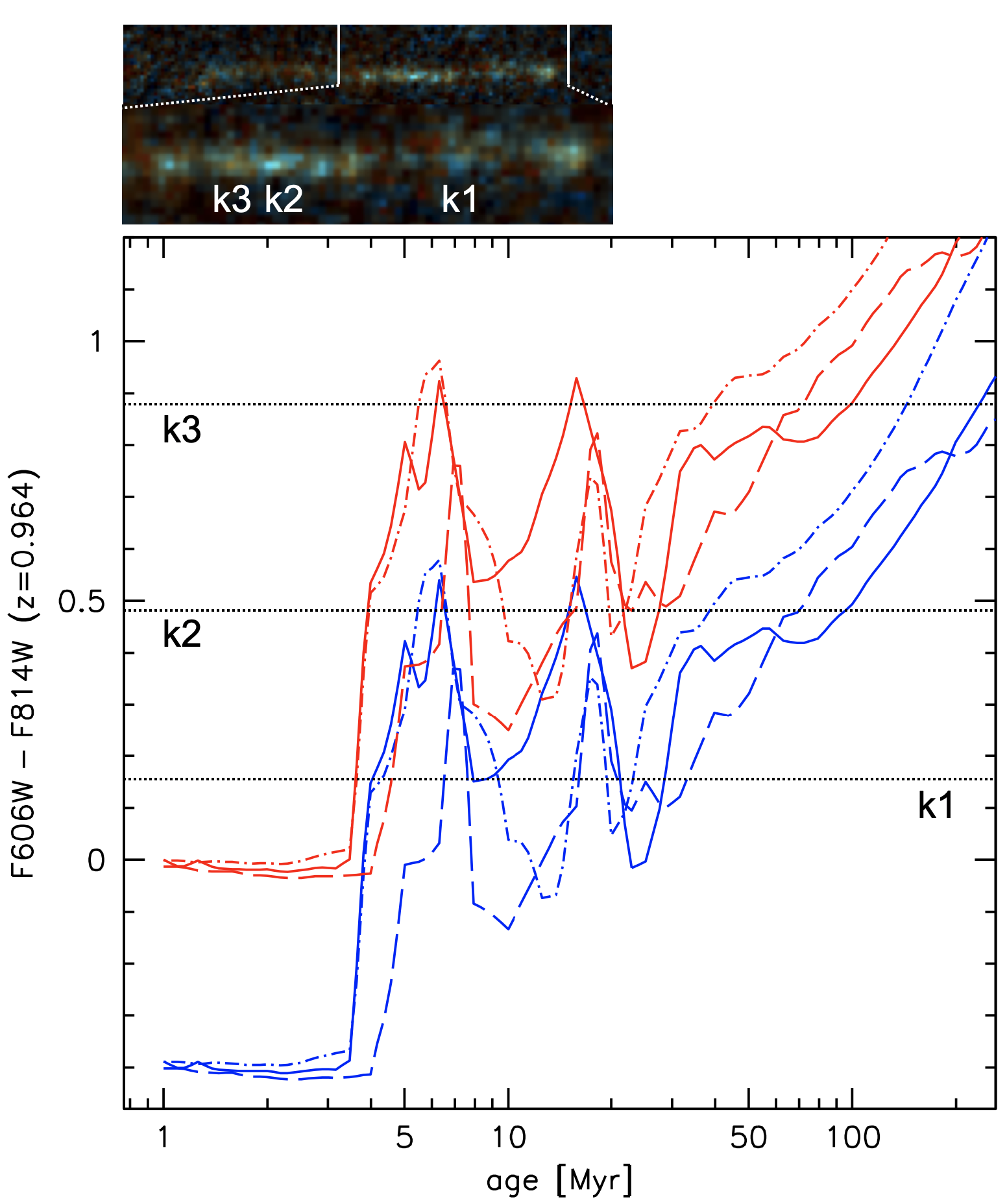}
  \end{center}
\vspace{-0.2cm}
    \caption{
Comparison of the observed colors of several
knots in the feature (shown at the top)
to model predictions
of \citet{conroy:09} for different ages. Dashed model
predictions are for a metallicty
$Z=-1$, solid for $Z=-0.5$, and dot-dashed lines are for $Z=0$. Blue
lines are dust-free models and red lines
illustrate the effect of dust attenuation
with $A_V=1$. Horizontal lines are measurements
for the three knots.
The ages of the youngest stars are likely $\lesssim 30$\,Myr,
but there is no straightforward relation between age and color in this
regime. The observed colors span a similar range as the models and
are consistent with a wide range of possible metallicities, ages,
and dust content.
}
\label{colors.fig}
\end{figure}

In between the two main shocks is a region where O stars
are probably the dominant source of ionization. 
At distances of $40$\,kpc\,$<r<$\,50\,kpc from the galaxy
the [O\,{\sc iii}]/H$\beta$ ratio is in the $1-2$ range and there are
several bright continuum knots.
These knots show strong ${\rm F606W}-{\rm F814W}$
color variation, mirroring the striking overall variation along
the feature that was seen in Fig.\ \ref{lineratios.fig}.
In Fig.\ \ref{colors.fig} we compare the measured colors of three
knots to predictions of stellar population
synthesis models. They were chosen
because they span most
of the observed color range along the feature.
The models span a metallicity range of $-1\leq Z\leq 0$
and have either no dust (blue) or $A_V=1$\,mag (red).  The metallicity
range encompasses that of the galaxy ($Z\approx -0.1$).

We find that the knots can indeed be young enough ($\lesssim 10$\,Myr)
to produce ionizing radiation. However, it is difficult to derive any
quantitative constraints
%The bluest of the three knots requires an age of $\lesssim 30$\,Myr, and could
%be as young as $\sim 1$\,Myr.
as there is no straightforward relation between age and color in
this regime. The reason for the complex model behavior
in Fig.\ \ref{colors.fig} is
that the ratio of red to blue supergiants changes rapidly
at very young ages \citep[``blue loops''; see, e.g.,][]{walmswell:15}.
%This 
%stars have lifetimes $<20$\,Myr \citep[see, e.g.,][]{chun:18} and
%are responsible for the red excursions 
%in the model curves of Fig.\ \ref{colors.fig} \citep{choi:16}.
We note that the evolution of supergiants is uncertain
\citep[see, e.g.,][]{chun:18}
and while the overall trends in the models are likely correct, the
detailed behavior at specific ages should be interpreted with caution
\citep[see, e.g.,][]{levesque:05,choi:16,eldridge:17}.
In \S\,\ref{interpretation.sec} we interpret the overall trend of
the color with position along the feature in the context of our proposed
model for the entire system.

Finally, we note that the knots appear to have a characteristic 
separation, as can be seen in Fig.\ \ref{colors.fig}
and in the pattern
of peaks and valleys from
$r=30$\,kpc to $r=50$\,kpc
in the F606W emission in Fig.\ \ref{lineratios.fig}.
The separation is $\approx 4$\,kpc. This could be coincidence or
be an imprint of a periodicity in the cooling cascade of the shocks.

\begin{figure*}[htbp]
  \begin{center}
  \includegraphics[width=1.0\linewidth]{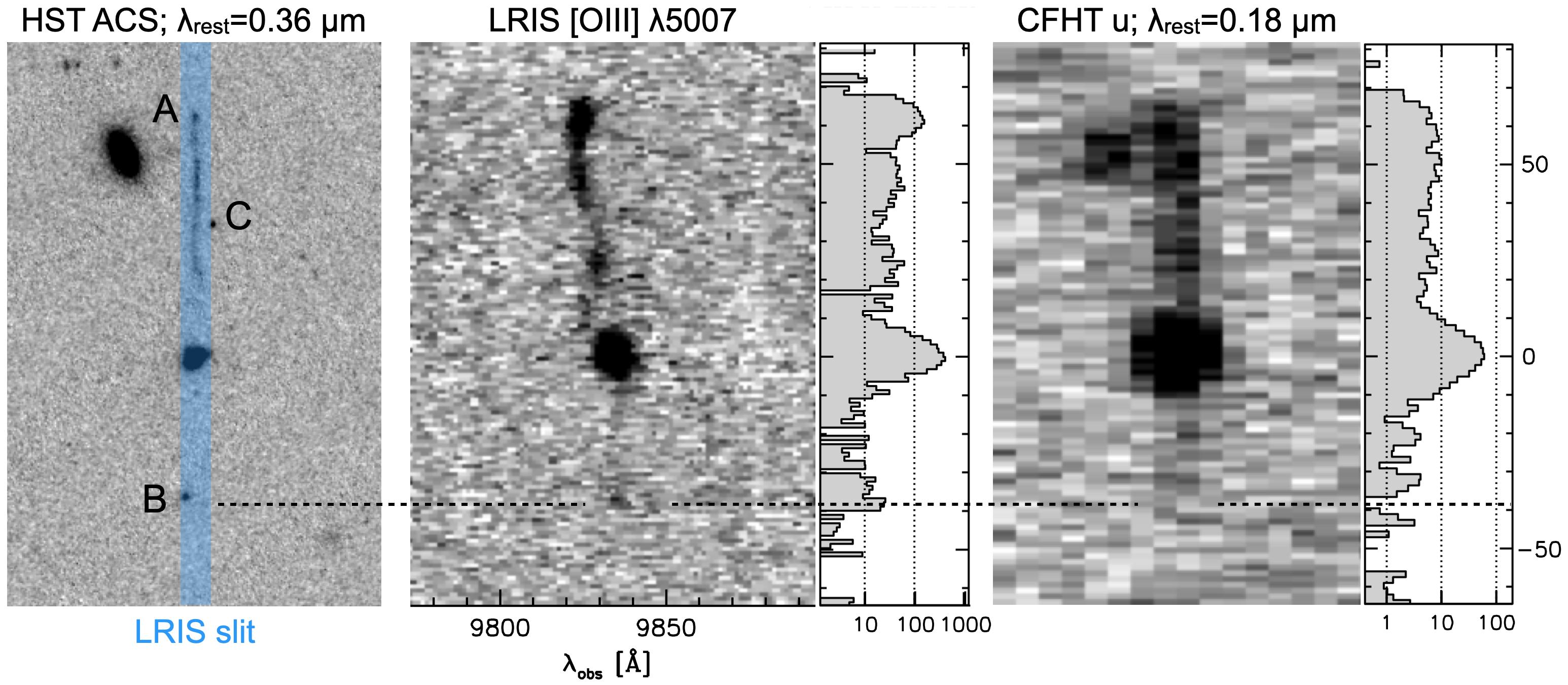}
  \end{center}
\vspace{-0.2cm}
    \caption{
{\em Left:} Section of the summed ACS F606W+F814W image, with the LRIS
slit indicated in blue. Besides the tip of the linear feature, A, 
there are two other bright spots in the vicinity, B and C. Object
B falls in the slit. {\em Center:} Section of the LRIS spectrum
around the [O\,{\sc iii}]\,$\lambda 5007$ line. Object B is detected,
as well as faint emission in between B and the galaxy. The attached
panel
shows the intensity along the feature, on a logarithmic scale.
{\rm Right:} The presence of a ``counter'' feature is confirmed through
its detection in the $u$-band, which samples the rest-frame far-UV.
For clarity the $u$-band image was
binned by a factor of 6 in the
direction perpendicular to the slit (and then expanded again to retain
the correct spatial scale).
Also note that the primary feature extends all the way to the galaxy,
in marked contrast to the pronounced gap between the galaxy and
the feature in the ACS image.
}
\label{abc.fig}
\end{figure*}

\section{A ``counter'' linear feature on the other side of the galaxy}
\label{counterwake.sec}

The LRIS slit covered the galaxy and the feature and
also extended beyond the galaxy on the other side. There is
no spatially-extended
F606W or F814W emission on this side
but there is an
unresolved object, ``B'', that is located at a distance
of $4\farcs 4$ from the galaxy within a few
degrees of the orientation of the feature (see Fig.\ \ref{abc.fig}).
The LRIS spectrum in the vicinity of the redshifted [O\,{\sc iii}] line
is shown in the middle panel of Fig.\ \ref{abc.fig}, after 
subtracting the continuum
and dividing by a noise model to reduce the visual effect
of sky residuals.

We detect a knot of [O\,{\sc iii}]\,$\lambda 5007$
emission near the location of B, redshifted by $\approx 40$\,\kms\ with
respect to the galaxy. Furthermore, there is evidence for
faint [O\,{\sc iii}] emission in between the galaxy and B.
This ``counter'' linear feature is also seen
in a $u$-band image, shown in the
right panel of Fig.\ \ref{abc.fig}. The object was serendipitously
observed with MegaPrime on the Canada France Hawaii
Telescope (CFHT) on September 11 and 12, 2020 in the context of
program 20BO44 (PI: A.\ Ferguson).
The total exposure time was 11,880\,s;
the data reduction is described in M.~L.~Buzzo et al., in
preparation.

The $u$-band surface brightness of the counter feature is
approximately $5\times$
fainter than on the other side, and it appears to terminate at the
location of the [O\,{\sc iii}] knot. Furthermore, the
primary feature extends all the way to the
galaxy in the $u$-band:
there is no gap at $r\lesssim 25$\,kpc as is the case in
the ACS data. The $u$-band
samples the rest-frame far-UV ($\lambda_{\rm rest} \approx
0.18\,\mu$m), and we conclude that the far-UV emission
of the entire system is largely decoupled from the near-UV
emission that is sampled with ACS.
The total far-UV brightness of the linear emission is
$\approx 70$\,\% of the far-UV brightness of the galaxy, whereas
this fraction is only
$\approx 40$\,\% at $\lambda_{\rm rest}\approx 0.36\,\mu$m.

The detection of the counter feature in the rest-frame far-UV
shows that the [O\,{\sc iii}] emission is likely real
and caused by shocks. The combination
of [O\,{\sc iii}] line emission and far-UV continuum emission
has been linked to cooling radiation
of fast ($\gtrsim 100$\,\kms) shocks, both theoretically
\citep[e.g.,][]{sutherland:93}, and observationally, for instance
in sections of supernova remnants \citep{fesen:21}.

It is difficult to determine the relationship between
object B and the counter feature. 
It has ${\rm F814W} = 25.28\pm 0.10$ (AB) and
${\rm F606W} - {\rm F814W} = 0.84 \pm 0.14$, and
it is misaligned by $4\arcdeg$ from
the line through A and the galaxy.
We will discuss the nature of B in the context of our preferred overall
model for the system in \S\,\ref{runaway.sec}.
%and \S\,\ref{conclusions.sec}.
%, that is, it has
%the same color as the galaxy albeit with considerable uncertainty.
%The implied mass is $\sim 3 \times 10^8$\,\msun.
There is also another compact object, C, that is nearly
exactly opposite to B in angle and distance. This object was not
covered by the LRIS slit and we have no information about it,
except that it is bluer than B. 
%It may be a faint
%star in the Milky Way halo, a compact low mass star-forming galaxy
%in the foreground or background, or an integral
%part of the A/B/C system.

\section{Interpretation}

\subsection{Various straight-line extragalactic objects}

With the basic observational results in hand we can consider possible
explanations. Thin, straight optical features that extend over several tens
of kpc have been seen before in a
variety of contexts. 
These include straight arcs, such as the one in
Abell 2390 \citep{pello:91}; one-sided tidal tails, with the Tadpole galaxy
(Arp 188) being the prototype \citep{tran:03b}; debris from
disrupted dwarf galaxies, like the multiple linear features
associated with NGC\,1097 \citep{amorisco:15};
ram pressure stripped gas, such as the spectacular 60\,kpc$\,\times\,1.5$\,kpc
H$\alpha$ feature associated with the Coma galaxy D100 \citep{cramer:19};
and ``superthin'' edge-on galaxies \citep{matthews:99}.

A gravitational lensing origin is ruled out by the identical
redshift of the galaxy that the feature points to. Tidal effects,
ram pressure stripping, or a superthin galaxy
might explain aspects of the main linear feature but are not
consistent with the shocked gas
and lack of rest-frame optical continuum emission on the other
side of the compact galaxy.
Given the linearity of the entire system, the
symmetry with respect to the nucleus, the presence of shocked
gas without continuum emission, as well as the brightness of
both the entire feature and the [O\,{\sc iii}] emission at the tip,
the most viable explanations all involve SMBHs -- either through nuclear
activity or the local action of a set of runaway SMBHs.

\subsection{An optical jet?}

Visually, the closest analog to the linear feature is the famous optical
jet of the $z=0.16$ quasar
3C\,273 \citep{oke:63,bahcall:95}: its physical size
is in the same regime (about half that of our object) and it
has a similar axis ratio and knotty appearance.
However, the detection
of bright emission lines along the feature is strong evidence against
this interpretation. The spectra of jets are power laws, and there are no
optical emission lines associated with optical jets or hot spots
\citep{keel:95}.

Furthermore, the
3C\,273 jet and 3C\,273 itself
are very bright in the radio and X-rays, with
different parts of the jet showing low- and high-energy emission
\citep[see][]{uchiyama:06}. 
We inspected the VLA Sky Survey \citep[VLASS;][]{lacy:20}
as well as a 60\,ks deep Chandra
image\footnote{https://doi.org/10.25574/05910} of the field that was obtained in 2005 in the context of program 5910 (PI Irwin). There is no evidence for
a detection of the linear feature or the
galaxy, with either the VLA or Chandra.
We note that the $z=0.96$ feature
might be expected to have an even higher X-ray luminosity
than 3C\,273 if it were a jet, as the contribution from Compton-scattered
CMB photons increases at higher redshifts \citep[see][]{sambruna:02}.

\subsection{Jet-induced star formation?}

Rather than seeing direct emission from a jet, we may be observing
jet-induced star formation \citep{rees:89,silk:13}.
There are two well-studied nearby
examples of jets triggering star formation,
Minkowski's object \citep{croft:06} and an area near a radio lobe of
Centaurus A \citep{mould:00,crockett:12}. There are
also several likely cases in the more distant Universe
\citep{bicknell:00,salome:15,zovaro:19}.
The overall idea is that
the jet shocks the gas, and if the gas is close to the
Jeans limit subsequent cooling can lead to gravitational collapse
and star formation \citep[see, e.g.,][]{fragile:17}. 
The presence
of both shocks and star formation along the feature is qualitatively
consistent with these arguments \citep[see][]{rees:89}.

The most obvious problem with this explanation is that
there is no evidence for nuclear activity in our
object from the BPT diagram, the VLASS, or Chandra imaging
(see above). It is possible,
however, that the AGN turned off between triggering star formation
and the epoch of observation, qualitatively similar to what is seen
in Hanny's Voorwerp and similar objects \citep{lintott:09,keel:12,smith:22}. 

A more serious issue is that
the morphology of the feature does not match simulations or
observations of jet-induced star formation. First, as can be
seen most clearly in the top right panel of Fig.\ \ref{overview.fig},
the feature is narrowest at the tip rather than the base. 
By contrast, for a constant opening angle a jet linearly increases
its diameter going outward from the host galaxy, reaching its greatest
width at the furthest point \citep[as illustrated by
HST images of the M87 jet, for instance;][]{biretta:99}.
Second, the interaction is most effective when the density
of the jet is lower
than that of the gas, and the shock that is caused by the jet-cloud interaction
then propagates largely {\em perpendicular} to the jet direction
\citep[e.g.,][]{ishibashi:12,silk:13,fragile:17}. This leads to
star formation in a broad cocoon rather than in 
the radial direction, as shown explicitly in the
numerical simulations of \citet{gaibler:12}. It is possible for
the jet to subsequently break out, but
generically jet-cloud interactions that are able to trigger
star formation will decollimate the jet.

A related problem is that the observed velocity dispersion of the shocked gas
is low. From the high resolution LRIS spectrum we find a velocity
dispersion of $\lesssim 20$\,\kms\  in the main shock at the tip of
the feature, which can be compared to $\sigma \sim 130$\,\kms\
in the shocked gas of Centaurus A \citep{graham:98}
and $\sigma \sim 50$\,\kms\ predicted
in recent simulations \citep{mandal:21}.
%Objects A and B at the ends of both linear features are also not
%easily accounted for in a jet-cloud interaction framework.
Most fundamentally, though, the feature is the inverse of what
is expected: the strongest interactions should not be at the furthest
point from the galaxy but close-in
where the ambient gas has the highest density, and the feature should
not become more collimated with distance but (much) less.

\begin{figure*}[htbp]
  \begin{center}
  \includegraphics[width=0.85\linewidth]{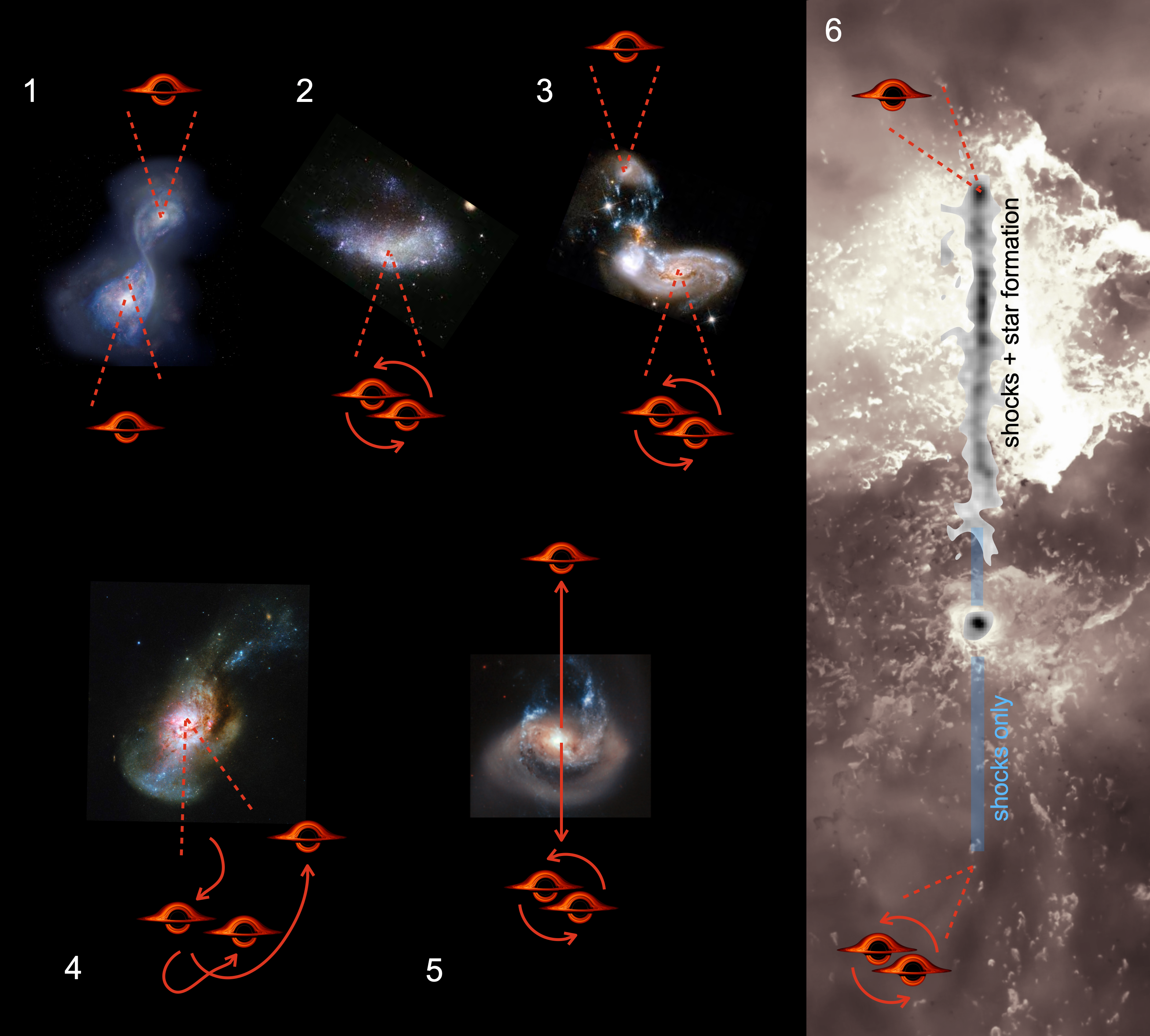}
  \end{center}
\vspace{-0.2cm}
    \caption{
Schematic illustration of the runaway SMBH scenario as an explanation of the key observed features. Panels 1--5 show a ``classical'' slingshot scenario \citep[e.g.,][]{saslaw:74}. First, a merger leads to the formation of a long-lived
binary SMBH (1,2). Then a third galaxy comes in (3), its SMBH sinks to the center of the new merger remnant, and this leads to a three-body interaction (4). One black hole (usually the lightest) becomes unbound from
the other two and receives a large velocity kick. Conservation of linear momentum implies that the remaining binary gets a smaller velocity kick in the opposite direction. If the kicks are large enough all SMBHs can leave the galaxy (5). There can be $\gtrsim 1$\,Gyr between the events in
panels (2) and (3). Panels (4) and (5) happened $\sim 40$\,Myr before the epoch of observation. The background of (6) is a frame from an Illustris TNG simulation \citep{tng}, with lighter regions having higher gas density. This illustrates that there can be highly asymmetric flows in the circumgalactic medium, and we speculate that the SMBH at A is traveling through such a region of relatively dense and cold CGM (see text).
}
\label{cartoon.fig}
\end{figure*}

\subsection{Runaway supermassive black holes}
\label{runaway.sec}

This brings us to our preferred explanation, the wake of a runaway
SMBH. The central argument is the clear narrow tip of the linear
feature, which marks both the brightest optical knot and the location
of very bright [O\,{\sc iii}] emission, combined with the apparent fanning out of
material behind it (as can be seen in
the top right panel of Fig.\ \ref{overview.fig}).
As discussed below (\S\,\ref{objectb.sec}) this scenario can
accommodate the feature on the other side of the galaxy,
as the wake of an escaped binary SMBH resulting from a three body interaction.
The properties of the (former) host galaxy can also be explained.
Its compactness and irregular
isophotes are evidence of the gas-rich recent merger  that
brought the black holes together, and
the apparent absence of an AGN reflects
the departure of all SMBHs from the nucleus.

\subsubsection{Mechanisms for producing the linear feature}

As discussed in \S\,1 there have not been many studies of the interaction
of a runaway SMBH with the circumgalactic gas, and there is no
widely agreed-upon description of what is expected to happen.
\citet{saslaw:72} focus on the direct interaction between gas that
is associated with the SMBH with the ambient gas. They predict a strong
bow shock which moves supersonically with the SMBH through the gas.
The aftermath of the shock leads to a cooling cascade, ultimately
leading to star formation in a wake behind the SMBH.
\citet{delafuente:08} study the gravitational effect of the
passage of a SMBH on the ambient gas. They find that small velocity kicks,
of up to several tens of \kms, are imparted on the gas, and that the
subsequent new equilibrium can lead to gravitational collapse and
star formation. There can be a delay between the passage of the SMBH
and the triggering of star formation, depending on the impact
parameter and the properties of the clouds.

Both mechanisms may be important; we certainly see evidence for both
star formation and shocks along the wake,
including potentially a bow shock at or just behind
the location of the SMBH itself, and conclude that the observations
are at least qualitatively consistent with the models that exist.
It is important to note that in these models
the star formation does not take place in
gas that was previously bound to the SMBH, but in the circumgalactic
medium. The kinematics and metallicity of the gas therefore
largely reflect its pre-existing state, perhaps slightly modified by
the passage of the SMBH.

\subsubsection{Nature of the counter wake}
\label{objectb.sec}

In this scenario there is 
only one explanation for the counter feature
on the other side of the galaxy (assuming it is real),
namely shocked gas in the
wake of a second runaway SMBH. This is not as far-fetched as it
may seem. When a third SMBH arrives in the vicinity of a pre-existing
binary SMBH, a common outcome of the three body
interaction is that one SMBH becomes unbound from
the other two. The post-interaction
binary can be the original one or contain the
new arrival \citep{saslaw:74}. 
In either case both the unbound SMBH and the binary
get a kick, in opposite directions and with the velocity inversely
proportional to the mass
\citep{saslaw:74,rees:75}. The counter feature is
then the wake of the most massive product of the three body
interaction, namely the
binary SMBH.

The relative projected
length of the wakes is 62\,kpc\,/\,36\,kpc\,=\,1.7:1. Here
we used the location of object B to determine the length of the counter
wake; using the location of the [O\,{\sc iii}] knot instead gives
the same ratio. Although
modified by their climb out of the potential well, this
length ratio is likely
not far from the velocity ratio of the black holes,
at least if $v_{\rm BH}\gg v_{\rm esc}$. Generally the least massive object is expected to escape (i.e., become
unbound) from the other two in a three-body interaction, with the
escape probability $\propto M_{\rm BH}^{-3}$ \citep{valtonen:91}. As the
escaped SMBH has a lower mass than each of the two components
of the binary, the velocity ratio between the single SMBH and the binary SMBH is then always $>2:1$, if linear momentum is
conserved.
A lower velocity ratio can work but only
if the three SMBHs all have
similar masses, for instance 4:4:3 for a:b1:b2, with
b1 and b2 the two components of the binary. In a 4:4:3
three body interaction
the probability that either one of the most massive objects
escapes (leading to the observed 1.7:1 ratio)
is about the same as the probability that the least massive one escapes.

We note that simulations
indicate that complete ejections of all SMBHs from the halo
are expected to be rare, occurring only in $\sim 1$\,\%
of three-body interactions \citep{hoffman:07}.
The dynamics are complex, however, particularly
when black hole spin, gravitational wave radiation, and gas flows
into the center are taken into account
\citep[see, e.g.,][]{escala:05,iwasawa:06,chitan:22}.
Along these lines,
a modification of the simple slingshot is that the binary hardens
due to the interaction with the third SMBH and merges,
leading to a gravitational recoil kick. This could explain how the binary made it so far out of the galaxy,
without the need for the three SMBHs to have near-equal masses.
However, the direction and amplitude of the recoil depends on the mass ratios, spins, and relative orientation of the binary at the time of the merger
\citep[e.g.,][]{herrmann:07,lousto:11}, and it seems unlikely that the two wakes would be  exactly opposite to one another in this scenario.

The counter wake is not only shorter than the
primary wake in the observed $u$-band but also much fainter, which
indicates that the shock has a lower velocity.
The shock (and black hole) velocities are undetermined -- although we will
constrain them in the next section -- but as noted above, the
velocity ratio between the wake and counterwake is likely $1.7$.
Assuming that the sound speed is similar on both
sides of the galaxy, the far-UV luminosity of fast shocks is expected to scale with the
velocity of the shock as $L_{\rm UV} \propto v_{\rm shock}^3$ \citep{dopita:95}.
The expected ratio of the UV surface brightness of
the two wakes is
therefore $1.7^3 = 5$, in excellent agreement with the observed ratio
(also 5; see \S\,\ref{counterwake.sec}). The post-shock pressure and
temperature scale as $\sim v_{\rm shock}^2$, and are therefore a factor of
$\sim 3$ lower in the counter wake. This may explain the lack of
gravitational collapse and star formation, although the local
conditions of the CGM may also play a role (see \S\,\ref{conclusions.sec}).

\subsubsection{Locations of the SMBHs}
\label{locations.sec}

The ``smoking gun'' evidence for this scenario would be the unambiguous
identification of the black holes themselves. 
The approximate expected (total) SMBH mass is
$M_{\rm BH} \sim
2\times 10^7$\,\msun, 
for a bulge mass of $7\times 10^9$\,\msun\ and assuming the
relation of \citet{schutte:19}.
The obvious places
to look for them
are A and B in Fig.\ \ref{abc.fig}. 
These are candidates
for  ``hyper compact stellar systems''
\citep{merritt:09}, SMBHs enveloped in stars and gas that escaped with
them. The expected sizes of HCSSs are far below the resolution limit
of HST and the expected stellar masses are bounded by
the SMBH mass, so of order $10^5$\,\msun\,--\,$10^7$\,\msun.

Focusing first on A, the tip of the feature is compact but not a point
source: as shown in the detail view of Fig.\ \ref{colors.fig} there are
several individual bright pixels with different colors embedded within
the tip. The approximate brightness of these individual knots
is ${\rm F814W}\approx 29.5$, after
subtracting the local background. This corresponds to a stellar mass
of $10^6$\,\msun\,--\,$10^7$\,\msun, in the right range for a HCSS.

The complex tip of the feature coincides with
very bright [O\,{\sc iii}] emission, and an interesting question is whether
this could be
the equivalent of the narrow line region (NLR) of an AGN. If so, it
is not composed of gas that is
bound to the black hole, as in that case the velocity dispersion
would be at least an order of magnitude higher. Instead, 
it would be a ``traveling'' NLR, with the accretion disk of the SMBH
illuminating the neighboring
circumgalactic medium as it moves through it. If the accretion
disk produces enough hard UV photons
to ionize the local CGM it should also emit X-rays.
The empirical relation between [O\,{\sc iii}]
luminosity and X-ray luminosity of \citet{ueda:15} implies
$L_{\rm X} \sim 3\times 10^{43}$\,ergs\,s$^{-1}$, and with standard
assumptions this correspond to $\sim 40$\,counts in the existing
60\,ks Chandra image. However, no object is detected, and
we tentatively conclude that it is unlikely that the SMBH at
A is active. This is not definitive and further
study is warranted: the \citet{ueda:15} relation
has significant scatter and
the object is on the edge of the
Chandra pointing, leading to a wide
PSF and relatively poor point source sensitivity.

We note that it is possible that the SMBH that is producing the
shocks and star formation at location A is not located there, but
is further than 62\,kpc from the galaxy. In the \citet{delafuente:08}
picture there is a delay between the gravitational impulse and 
the onset of star formation of about $\sim 30$\,Myr. For a black hole
velocity of $\sim 10^3$\,\kms\ this means that the SMBH may be several
tens of kpc ahead of the feature. A careful inspection of the HST image
shows no clear candidates for a HCSS beyond the tip.

Turning now to object B, it is a point source at HST/ACS resolution
that is clearly distinct from the shocked gas that
constitutes the counter wake. 
%The color of B is the same as that of the galaxy within
However, at ${\rm F814W}=25.3$ (see \S\,5)
it is uncomfortably bright in the context of
expectations for a HCSS.  
The stellar mass of B is 
$\sim 3\times 10^8$\,\msun\ if
the same $M/L$ ratio is assumed as for the galaxy,
an order of magnitude higher than
the probable black hole mass. 
%Fig.\ \ref{colors.fig},
%a color of ${\rm F606W}-{\rm F814W}=0.84$ is not only
%consistent with ages of $\gtrsim 50$\,Myr but could also indicate
%a dominant population of red supergiants in very young populations.
%Such young populations have much lower $M/L$ ratios, and 
%for an age of 5\,Myr we find a stellar mass of $M_{\rm B}
%\sim 2 \times 10^7$\,\msun.
%This just satisfies the constraint $M_{\rm HCSS}
%\lesssim M_{\rm BH}$; for a bulge mass of $7\times 10^9$\,\msun\ the
%expected SMBH mass is $M_{\rm BH} \sim
%2\times 10^7$\,\msun\ \citep{schutte:19}. We will return to this
%in \S\,\ref{conclusions.sec}.

\begin{figure}[htbp]
  \begin{center}
  \includegraphics[width=0.95\linewidth]{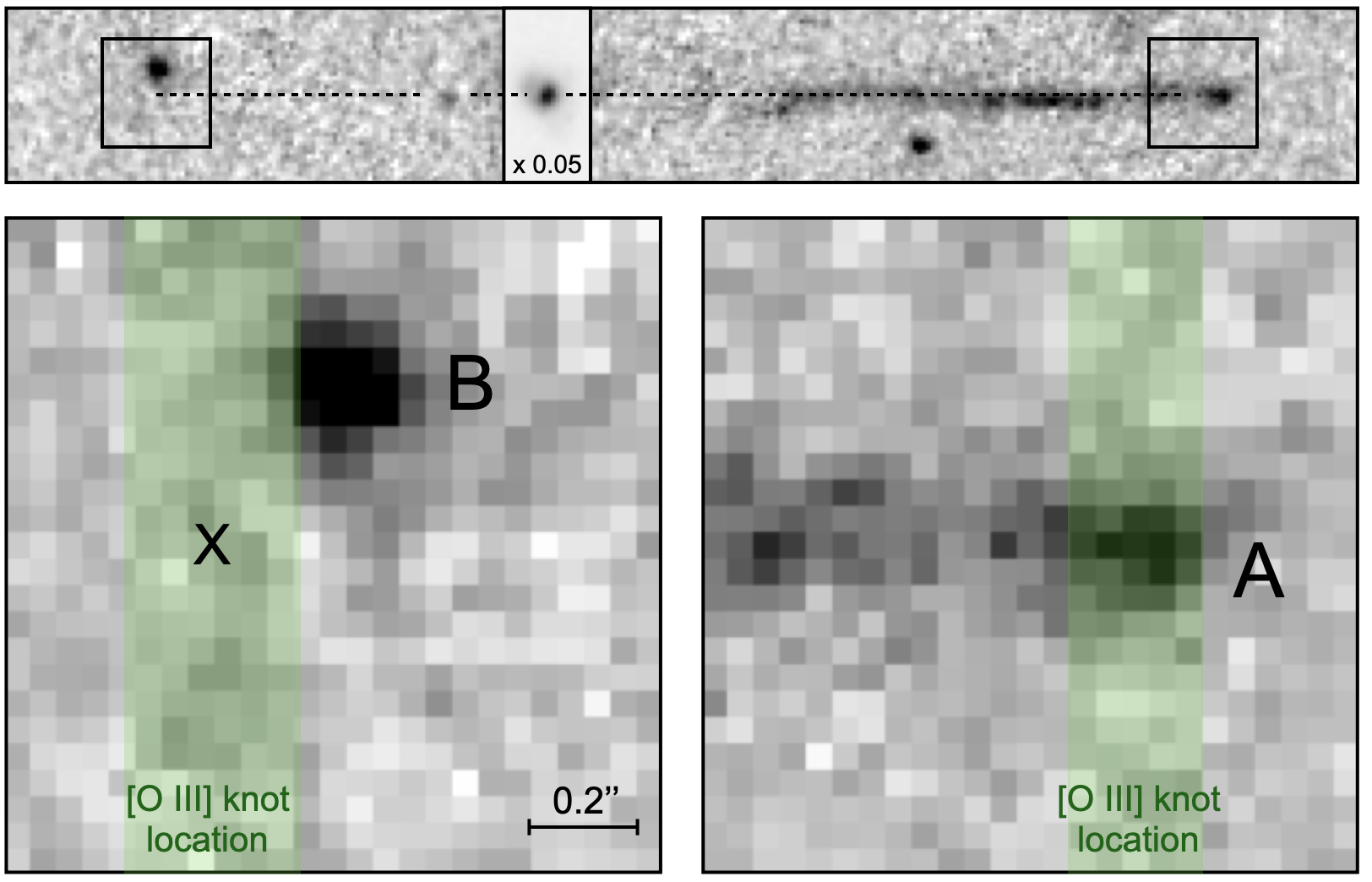}
  \end{center}
\vspace{-0.2cm}
    \caption{
Detailed view of the areas around A and B, in the summed
F606W\,+\,F814W image.
Green bands indicate the
locations of [O\,{\sc iii}] knots in the LRIS spectrum.
If
B is a chance projection along the line of sight, a
hyper compact stellar system may be detectable near the cross
in deeper data. In the vicinity of A,
the complex interplay of
shocks, star formation, and the SMBH itself could be
investigated with high resolution IFU spectroscopy. 
}
\label{ab.fig}
\end{figure}

A possible explanation for the brightness of
B is that it is a chance superposition of an unrelated object,
and that the apparent termination of
the counter wake at that location is coincidental. We show a detailed
view of the areas around A and B in Fig.\ \ref{ab.fig}. The green
bands indicate the locations of the [O\,{\sc iii}] knots on each
side of the galaxy, with the width of the band the approximate
uncertainty. The [O\,{\sc iii}] knot at the end of the counter wake
appears to be $0\farcs 25$ beyond B. Also, the angle
between B and the galaxy is $4\arcdeg$ offset from the angle between
A and the galaxy. There is no obvious candidate HCSS at the expected
location (marked by `X'), but that may be due to the limited depth
of the 1+1 orbit ACS data. 
%If there is
%no accretion after ejection, the expected mass of a HCSS for
%$v_{\rm BH}\sim 10^3$\,\kms\ and $M_{\rm BH}\sim 10^7$\,\msun\ is
%$\lesssim 10^5$\,\msun\ \citep{merritt:09},
%corresponding to ${\rm F606W}\gtrsim 30$.

Finally, object C is a third candidate HCSS, but only because of its
symmetric location with respect to B. In some dynamical
configurations it may be possible to split an equal-mass
binary, with B and C the two components, or to have multiple binary
black holes leading to a triple escape. These scenarios are
extremely interesting but also extremely
far-fetched, and without further observational evidence
we consider it most likely that C is a chance alignment
of an unrelated object.

\section{Modeling}

Here we assume that the runaway SMBH interpretation is correct, and
aim to interpret the details of the wake in the HST images in this
context. In \S\,\ref{interpretation.sec} we fit the seemingly random
color variations along the wake and in \S\,\ref{kinematics.sec}
we link the line-of-sight velocity variation along the wake to spatial
variations in the HST image. In both subsections
we assume that the SMBH is currently
located at position A and that it
triggered star formation instantaneously
as it moved through the circumgalactic gas.

\subsection{Stellar ages}
\label{interpretation.sec}

\begin{figure}[htbp]
  \begin{center}
  \includegraphics[width=0.95\linewidth]{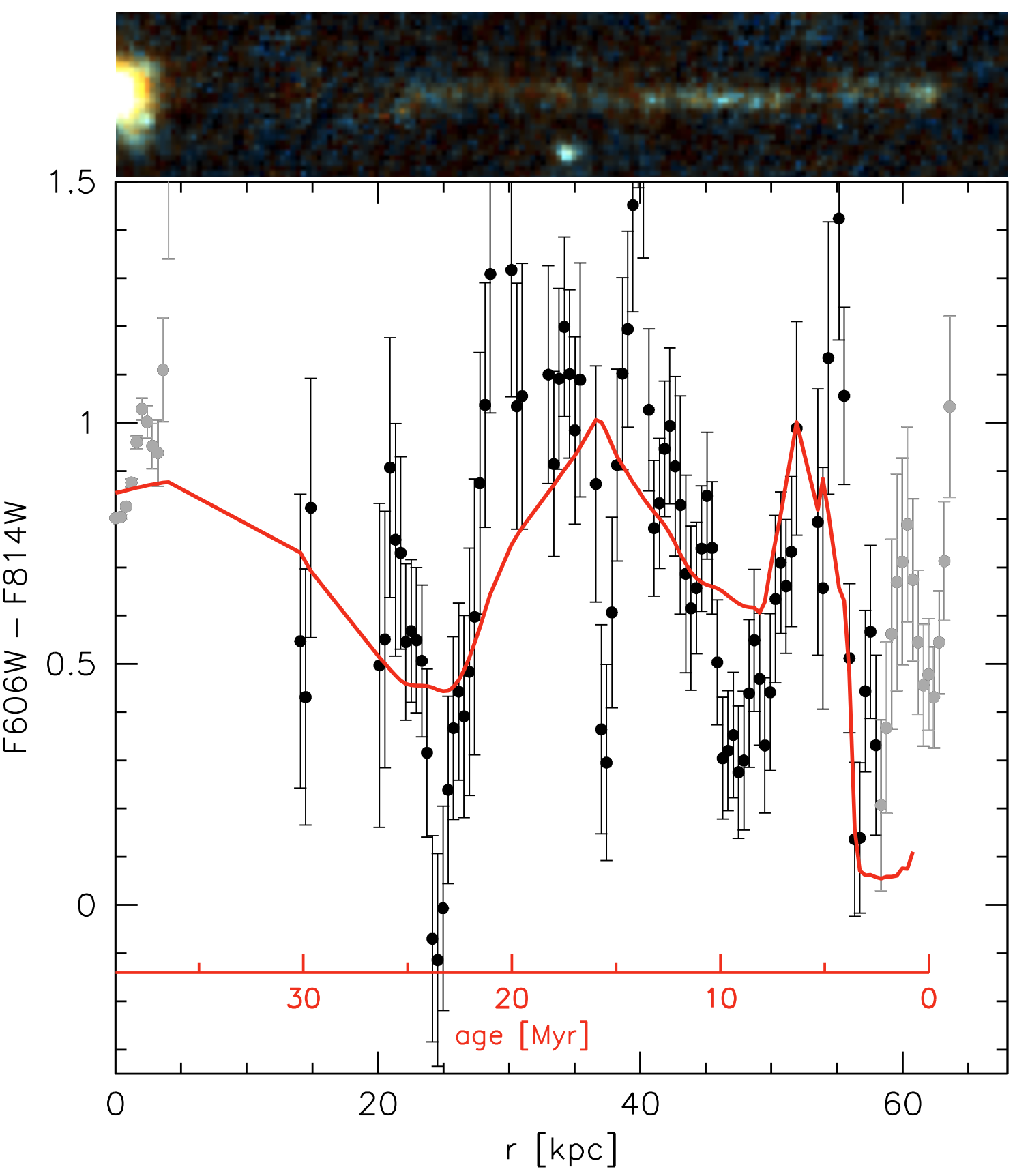}
  \end{center}
\vspace{-0.2cm}
    \caption{
Observed ${\rm F606W} - {\rm F814W}$ color along the wake,
after smoothing with a $0\farcs 15$ boxcar filter. The red curve
is a simple stellar population with $Z=-0.5$,
$A_V=1.1$\,mag, and age varying linearly with position
along the wake. The best-fit time since ejection is
39\,Myr, corresponding to a projected black hole velocity
of $v_{\rm BH}\approx 1600$\,\kms. 
}
\label{colors_age.fig}
\end{figure}

The color variation along the wake is shown in Fig.\ \ref{colors_age.fig}.
The information is identical to that in
Fig.\ \ref{lineratios.fig}, except we now show errorbars as well.
Colors were measured after averaging the F606W and F814W
images over $0\farcs 45$ (9 pixel) in the tangential direction and
smoothing the data with a $0\farcs 15$ (3 pixel) boxcar filter
in the radial direction. This
is why some prominent but small-scale features, such as the
blue pixel at $r=42$\,kpc, do not show up clearly in the color
profile. Data at $r>58$\,kpc are shown in grey as they
are assumed to be affected by the SMBH itself (the candidate
hyper compact stellar system ``A'' -- see \S\,\ref{locations.sec}).
Data at $r<5$\,kpc are part of the galaxy and not of the wake.

We fit the single burst stellar population synthesis models
of Fig.\ \ref{colors.fig} to the data.
The three metallicities shown in Fig.\ \ref{colors.fig}, $Z=0$,
$Z=-0.5$, and $Z=-1$, were fit separately. 
Besides the choice of metallicity there are two free parameters:
the overall dust content and the time since the SMBH
was ejected $\tau_{\rm eject}$. The age of the stellar
population $\tau'$ is converted to a position using
\begin{equation}
r' = 62 - 62\frac{\tau'}{\tau_{\rm eject}}.
\end{equation}
The best-fitting $Z=-0.5$ model has $A_V=1.1$ and
$\tau_{\rm eject}=39$\,Myr, and
is shown by the red curve in
Fig.\ \ref{colors_age.fig}. The other metallicities gave
similar best-fit parameters but much higher $\chi^2$ values.
This simple model reproduces the main color variation along the
wake, with three cycles going from blue to red colors
starting at $r=56$\,kpc all the way to $r=15$\,kpc.
As noted earlier, these large and sudden color changes
in the model curve
reflect the complex evolution of red and blue supergiants,
and are {\em not} due to a complex star formation history.
The red axis shows the corresponding age of the stellar population.

The best-fitting $\tau_{\rm eject}$
implies a projected black hole velocity of $v_{\rm BH}
\approx 1600$\,\kms. This velocity is in the expected range
for runaway SMBHs \citep[e.g.,][]{saslaw:74,volonteri:03,hoffman:07},
providing further evidence for this interpretation.
Specifically, it is too high for outflows and too low for relativistic
jets; besides hypervelocity stars, which are thought to have
a similar origin \citep{hills:88}, runaway SMBHs
are the only objects that are likely
to have velocities in this range.

\begin{figure*}[htbp]
  \begin{center}
  \includegraphics[width=0.97\linewidth]{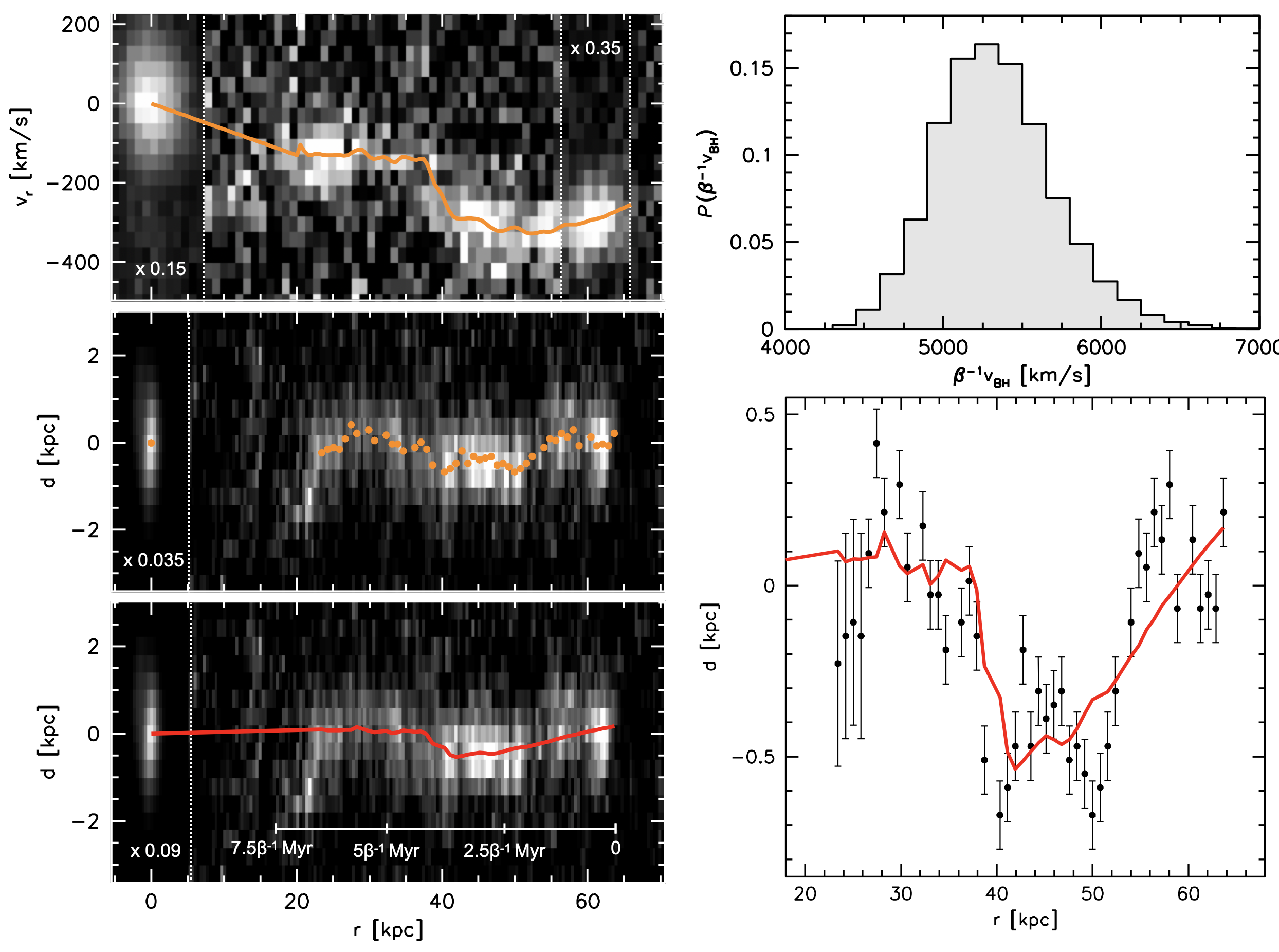}
  \end{center}
\vspace{-0.2cm}
    \caption{
Connection between velocities along the wake and its morphology.
{\em Top left:} [O\,{\sc iii}] emission along the wake, with a
fit to the velocity centroids in orange. {\em Middle left:}
HST image of the wake, with stretched vertical axis to emphasize
variations. The orange dots are centroids. {\em Bottom right:}
Fit of a kinematic model to the HST centroids,  based
on the [O\,{\sc iii}] velocity profile.
This fit is also shown in the bottom left panel. {\em Upper right:}
Distribution of posteriors for the  black hole velocity $v_{\rm BH}$,
modified by an
unconstrained geometric parameter $\beta$. For $\beta \approx 0.3$
we find that $v_{\rm BH}$ is consistent with the value derived
from the color variation along the wake.
}
\label{kinematics.fig}
\end{figure*}

\subsection{Kinematics}
\label{kinematics.sec}

The black hole velocity of $\approx 1600$\,\kms\ that we
derive above is much higher
than the observed line-of-sight velocities of gas along the wake, which
reach a maximum of $\approx 330$\,\kms\ (see Fig.\ 
\ref{overview.fig}). The observed velocities
reflect the kinematics of the circumgalactic medium: the passing
black hole triggers star formation in the CGM behind it but does not drag the
gas or the newly formed stars along with it.

In this picture the gas and newly formed stars will continue to move
after the black hole has passed. The wake should therefore not be
perfectly straight but be deflected, reflecting
the local kinematics of the CGM. We show the F606W\,+\,F814W HST image
of the wake in the middle left panel of Fig.\ \ref{kinematics.fig},
with the vertical axis stretched to emphasize deviations from linearity.
The wake is indeed not perfectly straight, but shows several
``wiggles'' with an amplitude of $\sim 0.5$\,kpc.
These deviations from a straight line are quantified by fitting
a Gaussian to the spatial profile at each position along
the wake and recording the centroids. These are indicated with orange
dots in the middle left panel and with black points with errorbars
in the bottom right panel.

The [O\,{\sc iii}]\,$\lambda5007$ velocity profile is shown in the top
left panel, with the orange line a spline fit to the changing
velocity centroids along the wake. The velocity profile shows a
pronounced change between 35\,kpc and 40\,kpc, where the line-of-sight
velocity increases from $\approx 150$\,\kms\ to $\approx 300$\,\kms.
There is a change at the same location in the spatial profile, suggesting
that the deviations from a straight line are indeed correlated with the
CGM motions.

We model the connection between the line-of-sight velocities and the wiggles
in the HST image in the following way. We assume that the
black hole leaves the galaxy in a straight line with velocity $v_{\rm BH}$
and that it triggers star formation instantaneously at each location
that it passes. The newly formed stars will move with a velocity $\beta
v_{\rm gas}$, where $v_{\rm gas}$ is the line-of-sight velocity measured
from the [O\,{\sc iii}] line and $\beta$ is a conversion factor between
line-of-sight velocity and velocity in the plane of the sky tangential to the wake.
By the time that the SMBH reaches $62$\,kpc, the stars at any
location along the wake $r$
will have moved a distance
\begin{equation}
\label{deq.eq}
d(r)=\beta v_{\rm gas}(r)\frac{62-r}{v_{\rm BH}}
\end{equation}
that is, the velocity in the plane of the sky multiplied by the time 
that has elapsed since
the passage of the black hole.

As $v_{\rm gas}$ is directly measured at all $r$, the only free parameter
in Eq.\ \ref{deq.eq} is $\beta^{-1} v_{\rm BH}$. In practice there
are several nuisance parameters: the model can be rotated freely with
respect to the center of the galaxy, and there may be an offset between the
line-of-sight velocity of the galaxy and that of the CGM at $r=0$. 
We use the {\tt emcee} package \citep{emcee}
to fit for the black hole velocity and
the nuisance parameters. The number of samples is 1200 with 300 walkers;
we verified that the fit converged.

The best fit is shown by the red line in the bottom right panel and
the bottom left panel of
Fig.\ \ref{kinematics.fig}. The fit reproduces the spatial variation
quite well, particularly when considering that $v_{\rm gas}$ is
measured from data with $8\times$ lower resolution. 
The posterior distribution of $\beta^{-1}v_{\rm BH}$ is shown
in the top right panel. We find $v_{\rm BH} =
\beta 5300^{+400}_{-300}$\,\kms. The constraint comes directly from
the amplitude of the wiggles: if the black hole velocity were lower
by a factor two, twice as much time would have passed since the
passage of the SMBH, and the wake would have drifted apart twice as
much ($\approx 1$\,kpc instead of the observed $\approx 0.5$\,kpc).

Combining this result with that from \S\,\ref{interpretation.sec} we
infer that the morphological deviations from a straight line
and the colors of the wake can be simultaneously
explained if $\beta \approx 0.3$,
that is, if the gas velocities perpendicular
to the wake are 30\,\% of the line-of-sight velocities. The implied direction
of motion is about 17$\arcdeg$ away from the line of sight
(with an unknown component in the plane of the sky along the wake).

\section{Discussion and Conclusions}
\label{conclusions.sec}

In this paper we report the discovery of a remarkable linear feature
that is associated with a galaxy at $z=0.96$. Although the
feature exhibits
superficial similarities to other thin objects, in particular the
optical jet of 3C\,273, close examination shows that it is quite unique
with no known analogs. 

We make the case that the feature is the wake of a runaway SMBH, relying
on the small number of papers that have been written on this topic
in the past fifty years \citep{saslaw:72,rees:75,delafuente:08}. This area
could benefit from further theoretical work, particularly since 
these papers propose a variety of formation mechanism for the wakes.
Hydrodynamical
simulations that model the shocks and
also take  gravitational effects
into account might bring these initial studies together
in a self-consistent framework.

Objects A and B are possible hyper compact stellar systems
\citep[HCSSs;][]{merritt:09}. Neither object is a clearcut case:
object A is not a point source, and the actual HCSS would be one of several
candidates within the main knot. Object B is brighter
than what might be expected for a HCSS
\citep[see][]{boylankolchin:04,merritt:09}, and as we show in \S\,6.4.3
it may well be a chance superposition of an unrelated object.
It could also be that \citet{merritt:09}  underestimate the mass
that can be bound to the black hole (as they do not take the effects
of gas or possible binarity of the SMBH into account), 
that the $M/L$ ratio of B is much lower than what
we estimate, 
%initial mass function is top-heavy \citep[as it appears
%to be in the Galactic center; see][]{lu:13}, 
or
that the SMBH is more massive than what we inferred from the
galaxy mass.

We show that the seemingly random color variation along the wake can
be explained by a simple model of aging of the stars,
beginning at the tip of the wake.
In this interpretation
the striking excursions in Fig.\
\ref{colors_age.fig} are due to the varying dominance of blue and red
supergiants.\footnote{We note that there is no appreciable contribution from emission lines in the HST filters; in particular, the redshifted
[O\,{\sc iii}] doublet falls redward of the long wavelength
cutoff of the F814W filter.} The evolution of these stars is quite uncertain;
turning the
argument around, the data provide a validation of the qualitative
behavior of the models from 1 to 30\,Myr.
The implied velocity of the SMBH at A is $v_{\rm BH}\sim 1600$\,\kms\ and the
velocity of the binary SMBH is
$v_{\rm BH}\sim 900$\,\kms\ if the ejection was symmetric.
These velocities are projected on the plane of the sky, and do not
correspond to predicted line-of-sight velocities;
the ratio between the line-of-sight velocities should
be $\sim 1.7$ but their absolute values are poorly constrained.

Velocities in this range are also indicated by the straightness of the
HST feature: as we show in \S\,7.2 the feature is expected to
differentially disperse, and its morphology requires that
it was created by a fast-moving object.
A third piece of evidence for high speeds comes from the
emission line ratios.
As noted in \S\,3.2 it is difficult to have [O\,{\sc iii}]/H$\beta$
ratios as high as $\sim 10$ unless there is a significant precursor 
component (photoionization ahead of the shock) and the shock has a velocity
of at least $\sim 500$\,\kms\  \citep{allen:08}.
We can speculate that the precursor component may be partially responsible
for the complexity of the tip of the feature: perhaps star formation
is not only triggered behind the SMBH but also just in front of it.

The shock velocity and luminosity provide a constraint on its
spatial extent. From Eqs.\ 3.4 and 4.4
in \citet{dopita:96} with
$L_{{\rm H}\beta}\sim 2\times 10^{40}$\,ergs\,s$^{-1}$ and
$v_{\rm shock} \sim 1600$\,\kms\ we obtain an
area of the shockfront
of $\sim 0.2n^{-1}$\,kpc$^2$, with $n$ the density in cm$^3$.
For $n<0.1$ (as expected for circumgalactic gas, even
with some gravitational compression)
the shock should be resolved at HST resolution,
and possibly even from the ground. In this context it
is interesting that there is some indication that
the [O\,{\sc iii}] emission is indeed resolved along the LRIS slit.
Turning this argument around,
a high resolution image of the shock  (in
either [O\,{\sc iii}] or the rest-frame far-UV) could provide
a joint constraint on the shock velocity and the density of the gas.

The measured line-of-sight velocities along the wake do not tell
us much about the  velocity of the SMBH and its accompanying shocks,
but they do provide a pencil beam
view of circumgalactic gas kinematics in a regime 
where we usually have very little information. We can compare the
kinematics to general expectations for halo gas.
The $z=1$ stellar mass -- halo mass relation implies a
halo mass of $\approx 3 \times 10^{11}$\,\msun\ \citep{girelli:20}
and a virial radius of $\approx 80$\,kpc \citep{coe:10}. 
Considering that the projected length of the wake is shorter than the
physical length, the $r_{\rm proj}=62$\,kpc wake
likely extends all the way to the virial radius.
Using $V_{\rm vir} = (GM_{\rm vir}/r_{\rm vir})^{0.5}$ we
have $V_{\rm vir} \approx 130$\,\kms, much lower than the
observed peak line-of-sight velocity of the gas of $\approx 330$\,\kms.
This difference may be due to the passage of the SMBH itself;
in the impulse approximation
of \citet{delafuente:08}, for example, the black hole imparts a velocity
kick on the ambient gas. An
intriguing alternative explanation is that the trajectory of the
SMBH intersected gas that is not in virial equilibrium but
an outflow or an inflow. An example of such a structure is
a cold stream that could be funneling gas toward
the galaxy. Such streams have been seen in simulations
\citep{keres:05,dekel:09}, although not yet observed.
A cold stream could explain why
the velocity dispersion of the gas is so low,
and perhaps also facilitated raising the density
above the threshold needed for gravitational collapse.
It might also explain why the line-of-sight velocity at the location of the
``counter'' [O\,{\sc iii}] knot,
on the other side of the galaxy, is much lower than the
velocities along the primary wake, and  perhaps also
why no star formation is taking place on
that side. We illustrate this possibility in the right panel of Fig.\
\ref{cartoon.fig}.

It is straightforward to improve upon the observations that are presented
here. The main spectrum is a 30\,min exposure with Keck/LRIS,
and the
exposure time for the
near-IR spectrum that was used to measure [N\,{\sc ii}]/H$\alpha$ 
was even shorter, 7.5\,min.
The
extraordinary sensitivity of the red channel of LRIS
enabled us to use the redshifted
[O\,{\sc iii}]\,$\lambda 5007$ line at $\lambda_{\rm obs}=9834$\,\AA\
for most of the analysis, despite the short exposure time.
Deeper data, for instance from the {\em JWST} NIRSPEC IFU, may show the expected
broad, highly red- or blueshifted emission lines of ionized gas that is bound
to the black holes themselves. Those data could also spatially resolve
flows, shocks, and star formation near A (see Fig.\ \ref{ab.fig}).
%\footnote{We see a hint of broad [O\,{\sc iii}] emission
%for object B in the 30\,min Keck spectrum, but deeper data are needed to
%verify this.}
The HST data is similarly shallow, at 1 orbit for each of the two ACS
filters. Deep ultraviolet imaging with UVIS is particularly
interesting, as that could map the spatial distribution of shocked gas
on both sides of the galaxy. A UVIS image would readily show whether
the counter wake is real, and whether it
points to B or is precisely opposite the main wake.
Finally, X-ray imaging could further constrain the
physics of the shock and the absorbing hydrogen column
\citep[see][]{dopita:96,wilson:99},
or even directly detect the accretion disk of one or more of the
SMBHs. The currently available 60\,ks Chandra image shows no
hint of a detection but as it is very far off-axis, there is
room for improvement.

Looking ahead, the morphology of the feature in the HST
images is so striking that it should
not be too difficult to find more examples, if they exist.
Future data from the Nancy Grace Roman telescope can be searched
with automated algorithms;
this is the kind of task that machine learning algorithms can be
trained
to do \citep[see, e.g.,][]{lochner:20}.
Although technically challenging, the most interesting
wavelength to search in
is probably the rest-frame far-UV, as it may
include cases where the SMBH did not trigger star formation.
Individual runaway SMBH systems are of great
interest in their own right; furthermore,
a census of escaped SMBHs can complement future gravitational
wave measurements
from LISA \citep{lisa} for a complete description of
SMBH evolution in -- and out of -- galaxy nuclei.

%can be tested: clear predictions
%however, can be confirmed but difficult to falsify

%future streaks striking morphology
%u band may be more common

%\begin{figure*}[htbp]
%  \begin{center}
%  \includegraphics[width=0.75\linewidth]{cartoon.png}
%  \end{center}
%\vspace{-0.2cm}
%    \caption{
%Illustration of the creation of a wake behind an escaped SMBH, after XX and
%YY.  Needs simulations.
%}
%\label{cartoon.fig}
%\end{figure*}

\vspace{0.5cm}
\noindent
%\begin{acknowledgements}
%We thank everybody.
We thank the anonymous referee for their constructive and helpful report.
Support from STScI grant HST-GO-16912 is gratefully acknowledged.
S.~D.\ is supported by NASA through Hubble Fellowship grant HST-HF2-51454.001-A.
%\end{acknowledgements}

%\software{
%\package{Astropy} \citep{astropy, astropy2018},
%\package{gala} \citep{gala},
%\package{numpy} \citep{numpy}
%}

\bibliography{bh}{}
\bibliographystyle{aasjournal}

\end{document}